\documentclass[12pt]{article}
\usepackage[top=3cm, bottom=3cm, left=3cm, right=3cm]{geometry}
\usepackage{amsmath}
\usepackage{amsfonts}
\usepackage{amssymb}
\usepackage{graphicx}
\graphicspath{{FiguresPDF/}}
\usepackage[colorlinks,citecolor=blue,linktoc=all,linkcolor=cyan]{hyperref}
\usepackage{graphicx}
\usepackage{svg}
\usepackage{mathtools}
\usepackage{upgreek}
\usepackage{physics}
\usepackage{multirow}
\usepackage{placeins} 
\usepackage{pgfplots}
\usepackage{subfigure}
\usepackage[singlelinecheck=false]{caption}
\usepackage{array}

\usepackage[T1]{fontenc}
\usepackage{dsfont}               
\usepackage{mathrsfs}             
\usepackage{slashed}              
\usepackage{amsbsy}
\usepackage{braket}
\usepackage{tikz}
\usepackage{bbold}

\providecommand{\keywords}[1]
{
  \small	
  \textbf{\textit{Keywords---}} #1
}

\usepackage[
    citestyle=numeric-comp,
    bibstyle=science,
    sorting=none,
    eprint=false,
    url=false,
    doi=true]{biblatex}
\DeclareSourcemap{
  \maps[datatype=bibtex]{
    \map{
      \step[fieldset=title, null]
    }
  }
}

\usepackage{xpatch}
\xpatchbibmacro{name:andothers}{ 
{\finalandcomma}%
}{%
\addspace%
}{}{}

\DeclareFieldFormat{doi}{%
  \mkbibacro{DOI}\addcolon\space
  \ifhyperref
    {\href{https://doi.org/#1}{\nolinkurl{#1}}}
    {\nolinkurl{#1}}}

\DeclareFieldFormat{labelnumberwidth}{\mkbibbrackets{#1}}

\AtEveryBibitem{\clearfield{month}}

\providecommand{\U}[1]{\protect\rule{.1in}{.1in}}

\pgfplotsset{compat=1.8}

\begin{document}


\title{Nonexponential decay law of the 2P-1S transition of the H-atom}

\author{F. Giacosa$^{1,2 *}$, K. Kyzioł$^{1 **}$
\\\\
$^1$ Institute of Physics, Jan Kochanowski University,\\ 
\textit{ul. Uniwersytecka 7, 25-406 Kielce, Poland.} \\$^{2}$ Institute for Theoretical Physics, J. W. Goethe University, \\ \textit{Max-von-Laue-Str. 1, 60438 Frankfurt, Germany.}}

\date{} 
\maketitle

\begin{abstract}
We evaluate numerically the survival probability $P(t)$ for the unstable 2P excited state of the hydrogen atom, which decays into the ground-state 1S emitting one photon ($\tau \sim 1.595$ ns), thus extending the analytic study of Facchi and Pascazio, Physics Letters A 241 (1998). To this end, we first determine the analytic expression of the spectral function of the unstable state, which allows for an accurate evaluation of $P(t)$.
As expected, for short and long times $P(t)$ shows deviations from the exponential law: a `Zeno' region occurs at extremely short times (up to $\sim 0.3$ attosec, followed by a longer `anti-Zeno' domain (up to $\sim 50$ attosec); at long times above $125 \tau$, the decay law scales as $t^{-4}$.
\end{abstract}


\begin{flushright}
*email: \href{mailto:fgiacosa@ujk.edu.pl}{fgiacosa@ujk.edu.pl} \\
**email: \href{mailto:kyziol11@wp.pl}{kyziol11@wp.pl}
\end{flushright}

\maketitle



\section{Introduction}

The fact that the decay law of a given unstable state, described  by the survival probability $P(t)$, is not simply an exponential function of the type $P(t) =e^{-t/\tau} =
e^{-\Gamma t}$, 
is well understood theoretically, e.g. Refs. \cite{LFonda_1978,Facchi:2008nrb,Kofman:2000gle}. In particular, deviations are expected at short and long times. At very short times, a quadratic decay law $P(t) = 1 -t^2/\tau_Z^2 + ...$ (where the coefficient $\tau_Z$ is the so-called Zeno time) is realized, which implies a larger survival probability than $e^{-t/\tau}$ and in turn renders the quantum Zeno effect possible \cite{Misra:1976by}: this is the slowing down of the decay if very frequent measurements are performed. Shortly after this Zeno region, an anti-Zeno domain that corresponds to a faster decay than $e^{-t/\tau}$ usually (but not necessarily) takes place: a~sequence of measurements at an appropriate time interval generates an increased decay rate. At very late times the decay follows a power law, $P(t) \sim t^{-\beta}$, where the exponent $\beta > 0$  depends on the specific system.
Quite interestingly, a very similar phenomenology applies also for relativistic decays which need a QFT treatment, see e.g. Refs. \cite{Facchi:1999ik,Giacosa:2011xa,Giacosa:2021hgl} (the particular case of strong decays in which large deviations are expected is discussed in Ref. \cite{Giacosa:2010br}). 

In general, these deviations occur at very short and at very long times, making them very difficult to observe in natural systems. 
On the experimental level, deviations from the exponential decay at short times (including both the Zeno and anti-Zeno domains and the related effects) were confirmed in Refs. \cite{1997Natur.387..575W,2001PhRvL..87d0402F} using an engineered tunneling of Na atoms through an optical potential. Deviations at long times were seen in fluorescence decays of chemical compounds in Ref. \cite{rothe}. Short- and late-time deviations were also confirmed by the analogous system of photons propagating in waveguide arrays \cite{crespi}. A type of `hidden evidence' of non-exponential decay law is reported for nuclear Beryllium decays in Ref. \cite{Kelkar:2004zz}.
Indeed, the previous examples show that, up to date, the deviations could be observed only in specific systems.

In this work, we intend to study a natural and very basic decay: the 2P-1S transition of the hydrogen atom. Namely, when the electron is located in the 2P orbital, it quickly ($\tau \sim 1.595$ ns) `jumps down' to the ground state emitting one photon. This decay represents then an optimal test to check what does it mean `short' and `long' times when deviations from the exponential decays are considered.
In Ref. \cite{Facchi:1998abc} this system has been studied by making use of analytic approximations. Here, we are able to determine  numerically $P(t)$ at a very good level of accuracy. This is possible because the spectral function of the unstable state is determined analytically. 

Our numerical results confirm in general the outcomes of Ref. \cite{Facchi:1998abc} but show also some novel aspects: (i) the numerical value of the Zeno-time $\tau_Z$ agrees very well with the outcome of Ref. \cite{Facchi:1998abc}, but the quadratic approximation is only valid for much shorter times (showing that the coefficient $\tau_Z$ is an upper limit); (ii) there is an anti-Zeno region that is much longer than the Zeno one; (iii) the late-time deviations start even later than the estimated value in Ref. \cite{Facchi:1998abc}.

The article is organized as follows: in Sec. 2 we recall some general features of the theoretical approach and show the spectral function of the 2P state; in Sec. 3 we present the main results of this work: $P(t)$ deviates form the exponential function at short and long times; finally, in Sec. 4 we discuss conclusions and outlooks. 

\section{The model and the 2P spectral function}

First, we briefly recall some main properties concerning the decay law, see
details in\ Ref. \cite{LFonda_1978}.\ The results can be also obtained in the specific
case of the Lee (or Lee-Friedrichs) Hamiltonian \cite{Facchi:2008nrb,Giacosa:2020tha,Giacosa:2022nyg}. This is a versatile model that implements a continuum of states and can be also extended to relativistic QFT cases \cite{Facchi:1999ik,Giacosa:2010br,Giacosa:2011xa,Giacosa:2021hgl,Zhou:2020vnz}.

Let us consider a system controlled by the Hamiltonian $H$ describing an
unstable system/particle $\left\vert S\right\rangle $ formed at the time
$t=0$. The survival probability $P(t)$ that the state has not decayed yet up to the
time $t>0$ is given by (see e.g. Ref. \cite{LFonda_1978}):
\begin{equation}
P(t)=|A(t)|^{2}\;\text{with }A(t)=\left\langle S\left\vert e^{-iHt}\right\vert
S\right\rangle \text{ ,}%
\end{equation}
where $A(t)$ is denoted as the survival probability amplitude. In turn, $A(t)$ can
be expressed as the Fourier transformation of the spectral function (also called energy
distribution) $d_{S}(E)$ of the unstable state: %

\begin{equation}
A(t)=\int_{E_{th}}^{\infty}\mathrm{d}E\,d_{S}(E)e^{-iEt}\;,
\label{amplitude}
\end{equation}
where $E_{th}$ is the lowest admissible `threshold' energy for the decay. 
The spectral function $d_{S}(E)$ emerges as the imaginary part of the
propagator $G_{S}(E)=[E-M+\Pi(E)]^{-1}$, where $M$ is the energy/mass of the unstable
state and $\Pi(E)$ the so-called self-energy (intuitively, describing processes
of the type $S\rightarrow$ decay product $\rightarrow S)$. It takes the
explicit form
\begin{equation}
d_{S}(E)=-\frac{1}{\pi}\operatorname{Im}[G_{S}(E)]=\frac{1}{\pi}%
\frac{\operatorname{Im}[\Pi(E)]}{(E-M+\operatorname{Re}[\Pi(E)])^{2}%
+(\operatorname{Im}[\Pi(E)])^{2}}\;,\label{spectral_function}%
\end{equation}
and is correctly normalized to unity, $\int_{E_{th}}^{\infty}\mathrm{d}%
E\,d_{S}(E)=1$.

Indeed, when the imaginary part $\operatorname{Im}[\Pi(E)]$ is known (or
modelled in some way), the real part can be obtained via the dispersion
relation
\begin{equation}
\operatorname{Re}[\Pi(E)]=\frac{1}{\pi}P\int_{-\infty}^{\infty}\mathrm{d}%
E^{\prime}\,\frac{\operatorname{Im}[\Pi(E^{\prime})]}{E^{\prime}-E}\,\text{.}%
\end{equation}
The decay width function $\Gamma(E)=2\operatorname{Im}[\Pi(E)]$ takes the
on-shell value $\Gamma(M)= \Gamma = 2\operatorname{Im}[\Pi(M)]=\tau^{-1},$ with $\tau$
being the lifetime of the unstable state.

\begin{figure}[!htb]
    \centering
    \subfigure{\includegraphics[scale=1]{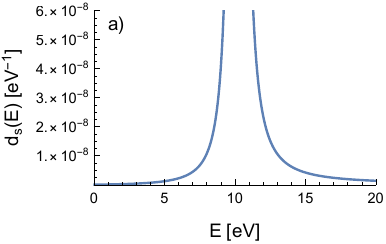}}
    \hfill
    \subfigure{\includegraphics[scale=1]{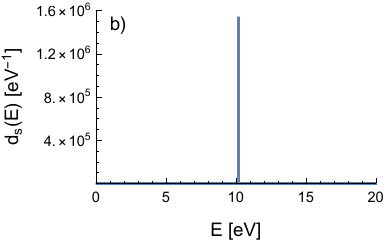}}
    \caption{Plot of the spectral function of the unstable state $S \equiv 2P$ for two different ranges of the vertical axis: as expected, the function is extremely narrow and peaked.}
    \label{plot_spectral_function}
\end{figure}

We can then move to the specific case of the 2P-1S transition.  It is
important to recall the explicit formula for the imaginary part of self-energy
function as presented in Ref. \cite{Facchi:1998abc} (see also the original calculations in Refs. \cite{PhysRevA.8.1710,SEKE1994269}):
\begin{equation}
\operatorname{Im}[\Pi(E)]=\pi\chi\Lambda\frac{\frac{E-E_{th}}{\Lambda}%
}{\Big(1+\big(\frac{E-E_{th}}{\Lambda}\big)^{2}\Big)^{4}}\,\vartheta
(E-E_{th})\;,\label{self_energy_function_imaginary_part}%
\end{equation}
where:
\begin{equation}
\chi=\frac{2}{\pi}\biggl(\frac{2}{3}\biggl)^{9}\alpha^{3}\simeq6.43509\times
10^{-9},\; \Lambda=\frac{3}{2}\,\alpha m_{e}\simeq5593.41\,\mathrm{eV}
\text{ .}
\end{equation}
Without loss of generality the threshold energy $E_{th}$ can be set to zero,
$E_{th}=0.$ Then, the energy of the state $2P$ (neglecting hyperfine splittings) reads
\begin{equation}
M=\frac{3}{8}\,\alpha^{2}m_{e}\simeq10.2043\,\mathrm{eV}\;.
\end{equation}
The on-shell physical decay width can be written down analytically as:
\begin{equation}
\Gamma = \frac{1}{\tau} = \frac{3}{2}\biggl(\frac{2}{3}\biggl)^{9} 
\frac{ m_{e} \alpha^{5} 
}{\Big(1+ \Big(\frac{\alpha}{4} \Big)^2 \Big)^{4}}\ 
 = 4.12582 \times 10^{-7} \, \text{eV ,}
 \label{ondw}%
\end{equation}
out of which 

\begin{equation}
    \label{lifetime}
    \tau \simeq 2.42376 \times 10^6 \; \mathrm{eV^{-1}} = 1.59535 \times 10^{-9} \, \mathrm{s} \text{ .}
\end{equation}
The real part of the loop $\operatorname{Re}[\Pi(E)]$ can be determined
analytically as: %

\begin{equation}
\begin{gathered} \frac{1}{\chi \Lambda}\operatorname{Re}[\Pi (E)]-C=-\frac{2\frac{E-E_{th}}{\Lambda}\ln \big( \frac{E-E_{th}}{\Lambda} \big)+\pi \big( \frac{E-E_{th}}{\Lambda} \big)^2}{2\Big( 1+\big( \frac{E-E_{th}}{\Lambda} \big)^2 \Big)^4}-\\-\frac{2\frac{E-E_{th}}{\Lambda}+\pi \big( \frac{E-E_{th}}{\Lambda} \big)^2}{4\Big( 1+\big( \frac{E-E_{th}}{\Lambda} \big)^2 \Big)^3}-\frac{4\frac{E-E_{th}}{\Lambda}+3\pi \big( \frac{E-E_{th}}{\Lambda} \big)^2}{16\Big( 1+\big( \frac{E-E_{th}}{\Lambda} \big)^2 \Big)^2}+\frac{15 \pi -16\frac{E-E_{th}}{\Lambda}}{96\Big( 1+\big( \frac{E-E_{th}}{\Lambda} \big)^2 \Big)} \; \text{ ,} \end{gathered}\label{self_energy_function_real_part}%
\end{equation}
where the subtraction constant $C$ is chosen such that $\operatorname{Re}[\Pi(M)]=0.$
The spectral function $d_{S}(E)$ can be determined by inserting results from
equations (\ref{self_energy_function_imaginary_part}) and
(\ref{self_energy_function_real_part}) into equation (\ref{spectral_function}).
Its form is shown in Fig. \ref{plot_spectral_function}: it is, as expected, an
extremely narrow function peaked at the energy $M$. 
We have numerically
verified that it is normalized to 1.

Once the shape of spectral function is fixed,  the decay law can be determined numerically.

\section{Results for $P(t)$}

\begin{figure}[!htb]
    \centering
    \includegraphics[scale=1]{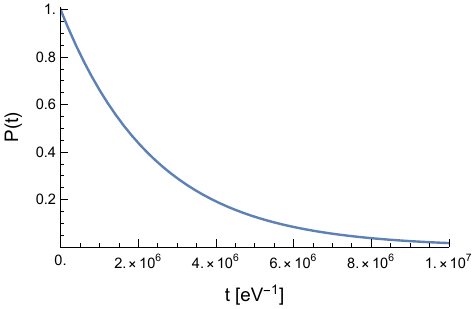}
    \caption{Survival probability at intermediate times (of the order of $\tau \sim 2.42 \times 10^{6}$ eV$^{-1}$). In this domain, the function is basically indistinguishable from the exponential decay.}
    \label{intermediate_times_plot}
\end{figure}

\begin{figure}[!htbp]
    \centering
    \subfigure{\includegraphics[scale=1]{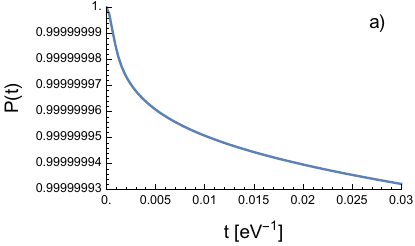}}
    \hfill
    \subfigure{\includegraphics[scale=1]{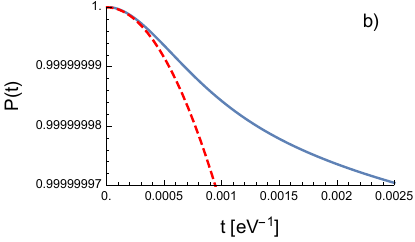}}
    \caption{Survival probability $P(t)$ for short times. In the left panel (a) we observe an anti-Zeno domain (enhanced decay rate). In the right panel (b) the Zeno domain is visible, as a direct comparison with Eq. (\ref{Zeno_time_definition}) shows.}
    \label{short_times_plot}
\end{figure}

At intermediate times (of the order of the lifetime $\tau$), the exponential decay law $P(t) = e^{-t/\tau}$ represents a very good approximation, see Fig.  \ref{intermediate_times_plot}.
At short times, by a direct numerical evaluation of Eq. (\ref{amplitude}), we obtain the results shown in Fig. \ref{short_times_plot}. For times of the order of $0.01$ eV$^{-1} 
\sim  10^{-18}$ s, deviations from the exponential law are visible, but this region is rather of the anti-Zeno type: an increased decay rate is realized (see below).

In order to make the quadratic Zeno-region visible, one needs to move to even shorter times of the order of $0.001$ eV$^{-1} \sim 0.6 \times 10^{-18}$ s. In this domain, the decay law can be expressed by
\begin{equation}
    \label{Zeno_time_definition}
    P(t)\simeq 1 - \frac{1}{2}\frac{\mathrm{d}^2P(t)}{\mathrm{d}t^2} \bigg|_{t=0} t^2 + ... = 1-\frac{t^2}{\tau^2_{Z}} + ... \text{ .}
\end{equation}
Namely, upon expanding the amplitude as 
\begin{equation}
    A(t) = 1 - it\braket{E} -\frac{t^2}{2}\braket{E^2} + ... \text{ ,} 
\end{equation}
one may easily see that $P'(0) = 0$ if $\braket{E}$ is finite. Moreover, the Zeno coefficient $\tau_Z$ reads

\begin{equation}
    \label{Zeno_time_result}
    \tau_Z = \sqrt{\frac{1}{\braket{E^2}-\braket{E}^2}}=\frac{1}{\sigma_E}\simeq 5.45911 \, \mathrm{eV^{-1}} = 3.59325 \times 10^{-15} \, \mathrm{s} \text{ .}
\end{equation}

\begin{figure}[!htbp]
    \centering
    \includegraphics[scale=1]{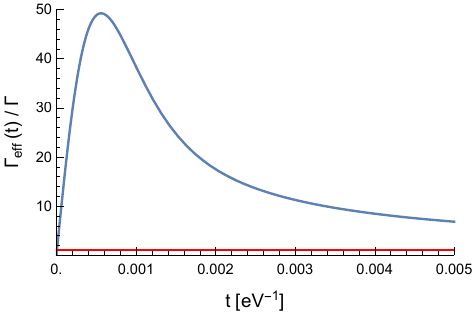}
    \caption{Ratio of the effective decay width over the exponential one: $\Gamma_{\mathrm{eff}}(t)/\Gamma$. The red curve corresponds to unity.}
    \label{effective_decay_width_plot}
\end{figure}

It is important to stress that in the present case, the Zeno time $\tau_Z$ is actually much longer than the non-exponential region in general and the quadratic region in particular. Namely, strictly speaking, the Zeno time defined in equation (\ref{Zeno_time_definition}), being the quadratic coefficient of Taylor expansion of $P(t)$, is not necessarily a good estimate of the non-exponential domain (but rather an upper limit of it).
To further investigate the decay rate at short times we introduce an effective decay width defined as \cite{Urbanowski1993,Urbanowski1994}:
\begin{equation}
    \label{effective_decay_width}
    \Gamma_{\mathrm{eff}}(t)= - \frac{\mathrm{d}P(t)}{\mathrm{d}t}\frac{1}{P(t)} \text { .}
\end{equation}
Namely, for a purely exponential decay $\Gamma_{\mathrm{eff}}(t)= \Gamma$ for each $t$. In turn, $\Gamma_{\mathrm{eff}}(t) > \Gamma$ signalizes an anti-Zeno domain, while $\Gamma_{\mathrm{eff}}(t) <  \Gamma$ a Zeno one. 
The function $\Gamma_{\mathrm{eff}}(t)/\Gamma$ is presented in Fig. \ref{effective_decay_width_plot}: it is clear that for short times an anti-Zeno region is present, with $\Gamma_{\mathrm{eff}}(t)/\Gamma$ having a maximum at
\begin{equation}
    t \simeq 0.00056 \, \mathrm{eV^{-1}} = 3.69232 \times 10^{-19} \, \mathrm{s} \text{ ,}
\end{equation}
at which the decay rate is about 50 times larger than the one in the exponential domain. The Zeno region takes place at such short times that it is barely visible in this figure. Note, quite similar, although non-identical, results for the decay width of electric-dipole transitions were obtained in Ref. \cite{PhysRevA.97.062122}.

We summarize in Table 1 relevant times concerning the anti-Zeno domain, which show that the anti-Zeno domain reduces to less than  $1 \%$  for times above 54 attosec.

\begin{table}
\caption{Selected numerical values of the effective decay width within the anti-Zeno domain together with the corresponding times.}
\centering
{\renewcommand{\arraystretch}{1.5}
\setlength\tabcolsep{20pt}
    \begin{tabular}{c|c|c}
     $\Gamma_{\mathrm{eff}}(t)/\Gamma$ & Time in $\mathrm{eV^{-1}}$ & Time in $\mathrm{s}$ \\
     \hline
     $2$ & $0.02130$ & $1.40183 \times 10^{-17}$ \\
   $1.1$ & $0.06242$ & $4.10857 \times 10^{-17}$ \\
  $1.01$ & $0.08234$ & $5.41941 \times 10^{-17}$ \\
    \end{tabular}}
\end{table}

\begin{figure}[!htbp]
    \centering
    \includegraphics[scale=1]{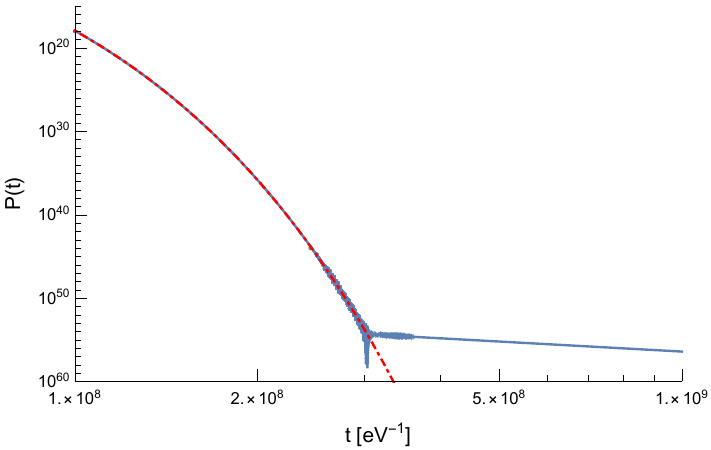}
    \caption{Survival probability at long times in log-log form. The red curve corresponds to purely exponential decay. An interesting feature is given by the fast oscillations close to the turn-over time.}
    \label{long_times_plot}
\end{figure}

Finally, we briefly describe the long-time domain. The pole position for the unstable 2P state is given by
\begin{equation}
    z_{pole} = M-i\frac{\Gamma}{2} = M-\frac{i}{2 \tau} 
    \text{ ,}
\end{equation}
out of which the survival probability amplitude can be obtained by changing the contour of the integration of Eq. (\ref{amplitude}) in the complex plane: one closes it between $(E_{th}=0,\infty)$ in the right-lower quadrant, the pole is picked up, and the vertical axis contribution must be subtracted, leading to:
\begin{equation}
    \label{survival_amplitude_new_contour}
    \begin{gathered}
        A(t)=-\frac{2i\operatorname{Im}[\Pi (z_{pole})] e^{-i z_{pole} t}}{z_{pole}-M+\operatorname{Re}[\Pi (z_{pole})]-i \operatorname{Im}[\Pi (z_{pole})]}-i\int_{0}^{\infty}\mathrm{d}y \, d_S(-iy) e^{-yt}
    \end{gathered}
    \text{ .}
\end{equation}
While for intermediate times the second term in Eq. (\ref{survival_amplitude_new_contour}) can be neglected, thus the decay is basically exponential, this is not true at long times, where the non-exponential part dominates.
Numerical calculations are typically difficult, so the optimal way to treat this problem is to approximate analytically the integral present in Eq. (\ref{survival_amplitude_new_contour}). 
At long times, only the linear term of the Taylor expansion of $d_S(-iz)$ matters, hence with good precision the survival amplitude can be approximated by the formula:
\begin{equation}
    \label{survival_amplitude_long_times}
    \begin{gathered}
        A(t)= - \frac{2i\operatorname{Im}[\Pi (z_{pole})]e^{-i z_{pole} t}}{z_{pole}-M+\operatorname{Re}[\Pi (z_{pole})]-i \operatorname{Im}[\Pi (z_{pole})]}-\frac{\chi}{M^2} \, t^{-2} 
    \end{gathered}
    \text{ .}
\end{equation}
The corresponding plot can be found in the log-log plot of Fig.  \ref{long_times_plot}, where the transition from exponential to power law is clearly visible. 

It is also possible to estimate the transition time at which the exponential law breaks down. Upon choosing the turn-over time as the one at which the absolute values of both terms present in expression (\ref{survival_amplitude_long_times}) are equal, one gets:
\begin{equation}
    t_{\text{turn-over}} \simeq 3.03297 \times 10^8 \, \mathrm{eV^{-1}} = 1.99634 \times 10^{-7} \, \mathrm{s} \simeq 125.1 \, \tau \text{ ,}
\end{equation}
basically implying that a detection of such long-time deviations is at present `de facto' impossible for the 2P-1S transition.

\section{Conclusions}

In this work, we have studied the non-exponential decay of a quite natural electromagnetic transition: the decay of an electron in the 2P state of an H-atom into the 1S level via the emission of a photon. 
While the quadratic Zeno region takes place at very short times ($\sim 0.3$ attosec $\simeq 1.88 \times 10^{-10} \, \tau$), the long-time deviations take place at very long times (0.2 $\mu$sec $\simeq 125 \, \tau$): both of them seem far from any experimental reach.

After a short Zeno region, there is a somewhat longer anti-Zeno domain (up to 50 attosec 
$\simeq 3.1 \times 10^{-8} \, \tau$). This is in agreement with the general discussion of Ref. \cite{Kofman:2000gle}, according to which the anti-Zeno domain (and effect) are in general easier to obtain. Indeed, in Ref. \cite{Giacosa:2019nbz} the anti-Zeno effect was proposed as an explanation of the neutron-decay anomaly \cite{Wietfeldt:2011suo}
due to an anti-Zeno region (and frequent measurements in the bottle-type experiments).

One may speculate that detection of the anti-Zeno effect (even if very difficult) could be possible by a continuous measurement \cite{1998PhRvA..57.1509S}
 of the ground state, in a set-up that would be analogous tho the optical experiment of Ref. \cite{2002OptCo.211..235B}. 
 Another interesting extension is the study of more decay channels, such as \cite{Giacosa:2021hgl,Giacosa:2019jxz}, as well as QFT relativistic systems.


\bigskip

\textbf{Acknowledgements:} This work was supported by the Minister of Science (Poland) under the `Regional Excellence Initiative' program (project no.: RID/SP/0015/2024/01).


\printbibliography

\end{document}